\documentclass[figures,preprint]{epl}
\usepackage{psfig}

\bibstyle{apsrev.bib}

\input epsf.sty

\title{Reentrance during nonequilibrium relaxation}
\shorttitle{Reentrance during nonequilibrium relaxation}

\author{L\'aszl\'o K\"ornyei\inst{1}, Michel Pleimling\inst{2}, and Ferenc Igl\'oi\inst{3,1}}
\institute{
\inst{1}Institute of Theoretical Physics,
Szeged University, H-6720 Szeged, Hungary\\
\inst{2}Institut f\"ur Theoretische Physik I, Universit\"at
Erlangen-N\"urnberg, D-91058 Erlangen, Germany\\
\inst{3}Research Institute for Solid State Physics and Optics,
H-1525 Budapest, P.O.Box 49, Hungary}

\begin{document}
\maketitle

\begin{abstract}
  We consider nonequilibrium critical dynamics of the two-dimensional
  Ising model for which the initial state is prepared by
  switching on random fields with zero mean and variance $H$. In the
  initial state there is no magnetic order but the clusters of
  parallel spins have a percolation transition for small enough $H$.
  Using heath-bath dynamics we measure the relaxation of the
  magnetization which shows a reentrance in time. Due to cluster
  dissolution in the early time regime there is a decrease of the
  magnetization, followed by an increase due to nonequilibrium
  domain growth which itself turns to a decrease due to equilibrium
  relaxation.  The power law decay of the nonequilibrium autocorrelation
  function is not influenced by the percolation in the initial state.

\end{abstract}

\newcommand{\bc}{\begin{center}}
\newcommand{\ec}{\end{center}}
\newcommand{\be}{\begin{equation}}
\newcommand{\ee}{\end{equation}}
\newcommand{\beqn}{\begin{eqnarray}}
\newcommand{\eeqn}{\end{eqnarray}}

\section{Introduction}

In nonequilibrium critical dynamics\cite{bray,godr,ritort,cg04} the system
under consideration is prepared in some initial state from which it is
quenched to the critical temperature, $T_c$, and then let to evolve in time
according to the given dynamical rules. Generally one is interested in
the relaxation of the magnetization, $m(t)=\langle \sigma(t)\rangle$,
and in the behavior of the autocorrelation function, $G(t,s)=\langle
\sigma(t) \sigma(s)\rangle$. Here $\sigma(t)$ denotes the operator of
the order parameter and $s$ and $t$ are the waiting time and the
observation time, respectively.

In most of the studied cases the initial state is of two kinds: it is
either the (completely) ordered one, or the (completely) disordered
one. Starting from the ordered state with the initial 
temperature $T_i=0$, the critical
relaxation process involves only equilibrium critical exponents.
For example the decay of the magnetization is asymptotically given by
\be
m(t) \simeq t^{-x/z},\quad T_i=0\;,
\label{m_T0}
\ee
where $x=\beta/\nu$ is the
anomalous dimension of the magnetization and $z$ is the dynamical
exponent. The decay of the autocorrelation function follows the rule:
\be
G(t,s) \sim (t-s)^{-2x/z} g(t/s),\quad T_i=0\;,
\label{G_T0}
\ee
Where the scaling function $g(y) \sim y^{x/z}$ for $y \gg 1$ and
tends to a constant otherwise. This means that $G(t,s)$ decreases as $t^{-x/z}$
for $s \ll t$, similar to the magnetization in Eq. (\ref{m_T0}), whereas for
$t/s=O(1)$ the decay is the same as for the equilibrium
autocorrelation function.

If on the contrary the initial state is the completely disordered one
with $T_i \to \infty$, new type of
nonequilibrium exponents appear in the relaxation process. Starting with a small
initial value, $m_i$, the magnetization behaves as\cite{jss89}
\be
m(t) \sim m_i t^{\theta}, \quad t < t_i,\quad T_i=\infty\;.
\label{m_Tinf}
\ee
Here $\theta$ is the initial slip exponent, which is zero in
mean-field theory but generally positive in realistic systems.
Consequently in the initial period, $t<t_i$, 
the order in the system is increasing due to critical
fluctuations. The limiting
time-scale $t_i$ can be obtained from the condition that for $t>t_i$
the relaxation goes over to the decay of Eq. (\ref{m_T0}), leading to
the estimate:
\be
t_i \sim m_i^{-z/x_i},\quad x_i=\theta z+x\;.
\label{t_i}
\ee
Here $x_i$ is called the anomalous dimension of the initial
magnetization. Note that $t_i$ tends to infinity as $m_i$ goes to
zero.  Another new feature of nonequilibrium dynamics starting from a
disordered initial state is given by the fact that the autocorrelation function, which is
measured in the state with $m_i=0$, is non-stationary: $G(t,s)$
depends both on $t$ and $s$ and not only on the time difference 
$t-s$. In the limit $t \gg s$ the decay is given by\cite{huse89}:
\be
G(t,s) \sim t^{-\lambda/z},\quad t \gg s,\quad T=\infty\;,
\label{G_Tinf}
\ee
and the autocorrelation exponent $\lambda$
satisfies the scaling relation\cite{jans92} $\lambda=d-\theta z=d-x_i+x$.

At this point we can note a formal analogy between nonequilibrium
critical dynamics and static critical behavior in semi-infinite
systems\cite{binder,diehl,ipt,pleim}. In the former problem translational invariance is broken in
time due to the quench, whereas in the latter phenomenon translational
invariance is broken in space in the direction perpendicular to the
surface of the system. The initial state in dynamics with $T_i=0$
($T_i=\infty$) corresponds to a semi-infinite system with fixed (open)
boundary conditions. In this respect the exponent $x_i$ is analogous
to the surface magnetization exponent $x_1$ at an ordinary surface
transition.

In nonequilibrium critical dynamics there are a few examples in which
the initial state contains nontrivial power-law correlations. Here we mention relaxation in
the 2d XY-model\cite{Berthier,abriet} when the initial state is prepared at a temperature $T_1>0$ and
the quench is made to a temperature $T_2>T_1$ with both temperatures in
the critical phase, i.e.\ $T_2<T_{KT}$, where $T_{KT}$ is the
Kosterlitz-Thouless temperature. According to spin-wave theory for
$T_2 \ll T_{KT}$ the autocorrelation function in Eq. (\ref{G_T0}) is modified by
replacing $x$ by $x(T_2)-x(T_1)$, where $x(T)$ is the temperature dependent
anomalous dimension. This theoretical prediction has been checked numerically\cite{abriet1}.
In another work the nonequilibrium dynamics of the
d-dimensional spherical model is studied starting from an initial state which
has long-range correlations\cite{picone}. Depending on $d$ and the correlation exponent $\alpha$,
different types of critical ageing behavior are found.
In both examples described above there is
quasi-long-range order in the initial state which is characterized by a fractal dimension
$d_f^i$ larger than the fractal dimension $d_f^c=d-x$ of the critical state.

In the present paper we consider a type of initial states in which
the disordering effect is not due to temperature, but is due to switching on
random fields of zero mean and variance $H$. In the limiting case of
strong fields, $H \gg J$, $J$ being the coupling of the system, we
recover the completely disordered initial state with $T_i=\infty$. In
the other limiting case, $H \to 0$, the initial state is completely
ordered as for $T_i=0$.  We are interested in the behavior of
nonequilibrium dynamics for intermediate values of $\Delta=H/J=O(1)$.
To be concrete we consider the two-dimensional Ising model for which
random fields of any strength destroy the magnetic long-range order\cite{aizenman}.
However the structure of the initial state on the level of clusters of
parallel spins shows interesting features\cite{seppala}. For strong enough fields,
$\Delta>\Delta_{perc}$, these clusters have a finite extent, whereas
for weaker fields, below $\Delta_{perc}$, there is a percolation
phenomenon: there are giant clusters for both spin directions which
percolate the sample. According to numerical results\cite{seppala,kornyei} these clusters
are isomorph with those of short range percolation having the same
fractal dimension\cite{staufferaharony} $d_f^{perc}=91/48$.

In this paper we want to clarify
the effect of the change in the initial state topology on the properties of
the nonequilibrium dynamics. In particular we want to see how the percolation of parallel
spins is manifested in the magnetization relaxation. 

\section{Numerical results} 

The initial state used in our nonequilibrium relaxation measurements is generated by switching on
random fields, so that the Hamiltonian of the system is defined by:
\be
H=-J \sum_{\langle ij \rangle} \sigma_i \sigma_j -\sum_i h_i \sigma_i\;,
\label{HRFIM}
\ee
where $\sigma_i=\pm 1$ is an Ising spin located at site $i$ of a square lattice. The
nearest-neighbor coupling, $J>0$, is ferromagnetic whereas the $h_i$ random
fields are taken from a symmetric Gaussian distribution:
\be
P(h_i)=\frac{1}{\sqrt{2 \pi H^2}} \exp\left[-\frac{h_i^2}{2 H^2} \right]\;,
\label{PG}
\ee
which has a variance $H$. The ground state of the system for a given
realization of the random fields is exactly calculated by a
combinatorial optimization method\cite{heiko} which works in strongly
polynomial time. With help of this effective algorithm systems as
large as $L=500$ are treated numerically and averaging is performed
over at least 80000 realizations.

As we have already mentioned in the introduction the magnetic
correlation function in the initial state, $\langle \sigma_i \sigma_j
\rangle$, is short-ranged, so that the magnetic correlation length
$\xi_m$ is finite. However, considering the topology of clusters of
parallel spins other length-scales can be defined, which can diverge
by varying the parameter $\Delta=H/J$. One of these lengths is the
so-called breaking-up length, $l_b$, which measures the typical
fluctuations of the interface separating the different spin
orientations. In other words in a finite system with $L<l_b$ the
ground state is homogeneous, but for $L>l_b$ both
spin orientations are present in the ground state and one can use a coarse-grained
description with effective cells of size $\sim l_b$.  For weak random
fields the breaking-up length asymptotically behaves as\cite{lb}
\be
l_b \sim \exp(A/\Delta^2)\;,
\label{l_b}
\ee
thus it is divergent as $\Delta \to 0$.  The other length appearing
in the geometry of the initial state is the percolation correlation
length, $\xi_{perc}$, which is the measure of the typical size of the
largest clusters in the system. $\xi_{perc}$ is divergent below the
percolation transition point, which is estimated as\cite{seppala} $\Delta_{perc}
\approx 1.65$.

In the numerical calculations we considered the disorder in such a
range for which the coarse-grained description is applicable, i.e. the
breaking-up length satisfies the relation: $\xi,L \gg l_b(\Delta)$. In
this way we made the calculations for $1.25\le \Delta \le 4$. For example, for
$\Delta=1.4$, deeply in the percolation regime, the breaking-up length is $l_b \approx 25$ \cite{seppala}.

\begin{figure}
\centerline{\epsfxsize=3.1in\epsfbox
{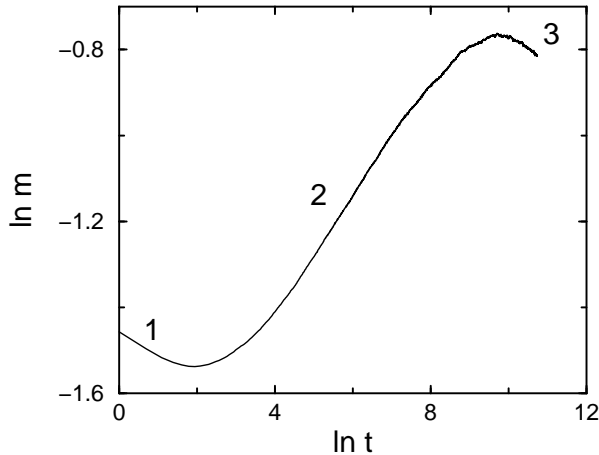}
}
\caption{Typical temporal evolution of the magnetization in a critical relaxation measurement when starting 
from a RFIM ground state. In order to make the three different regimes and the reentrance in time
clearly visible, we show here data obtained for a large initial magnetization $m_i=0.25$. The variance of the
field is $\Delta=1.7$ und the system contains $200 \times 200$ spins.
\label{fig1}}
\end{figure}
\begin{figure}
\centerline{\epsfxsize=4.1in\epsfbox
{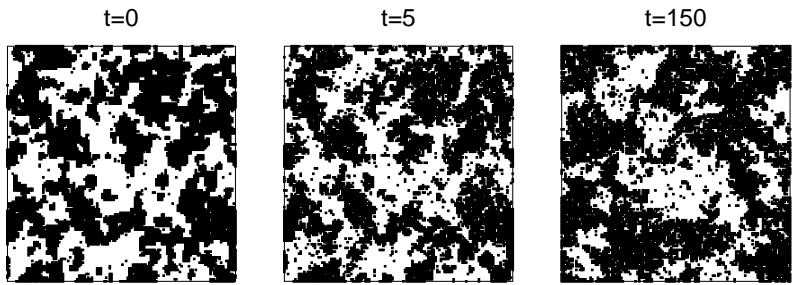}
}
\caption{Snapshots of the spin configuration of a system with $128 \times 128$ spins and $m_i = 0.0390625$
at three different times:
ground state at t=0 with $\Delta = 1.7$, close to the minimum at $t=5$, and in the increasing part at $t=150$.
\label{fig2}}
\end{figure}

Having prepared the system in the initial state with a small magnetization $m_i$, we quench
it to the critical point and study the nonequilibrium relaxation process, thereby
using the standard heat-bath algorithm.
As shown in Figure \ref{fig1} the magnetization displays an intriguing reentrance in time for not too large
values of $\Delta$: immediately
after the quench the magnetization first decreases (regime 1), reaches a minimum at time $t_{min}$ (the value of
which depends on $\Delta$ and $m_i$), increases in the regime 2  with $t_{min} < t < t_{max}$, 
before decreasing again in the large time
limit with $t > t_{max}$ (regime 3). This non-monotoneous behaviour of the magnetization is understood by looking at the
snapshots of Figure \ref{fig2}. At $t=0$ the system is formed by homogeneous cells wherein the spins have the
same orientation. The typical linear size of these cells is $l_b$. Immediately after the quench these compact cells
are dissolved which leads to a decrease of the magnetization (see the snapshot
for $t=5$). As this process is exactly the same as when starting 
from a totally ordered initial state (corresponding to one large, the system filling, homogeneous cell)
the decay of the magnetization follows the power law (\ref{m_T0}). This dissolution stops when the total magnetic 
moment of the cells is of the order $O(1)$, i.e.\ when the magnetization has been reduced by a factor $l_b^d$ where
$d=2$ is the dimensionality of the system. This yields the prediction 
\begin{equation} \label{tmin1}
t_{min}^{-x/z} \sim l_b^d
\end{equation}
and therefore
\begin{equation} \label{tmin2}
\ln t_{min} \sim \ln l_b \sim \frac{1}{\Delta^2}.
\end{equation}
After the cells have been dissolved, the usual nonequilibrium domain growth sets in yielding an increasing
magnetization in the regime 2. This increase of order (see the snapshot in Figure  \ref{fig2} at $t=150$) is
due to the critical fluctuations and is therefore similar to that observed when starting from a completely
disordered initial state with a small magnetization. We therefore expect that the temporal evolution
of the magnetization in the regime 2 is also governed by the initial slip exponent $\theta$, see Eq. (\ref{m_Tinf}),
and that the domain growth continues up to a time $t_{max}$ with $\ln t_{max} \sim \ln t_{min} + \ln t_i$
where $t_i$ is given in Eq.\ (\ref{t_i}). For longer times
(regime 3) the relaxation again goes over to the well-known decay (\ref{m_T0}).

\begin{figure}
\centerline{\epsfxsize=3.1in\epsfbox
{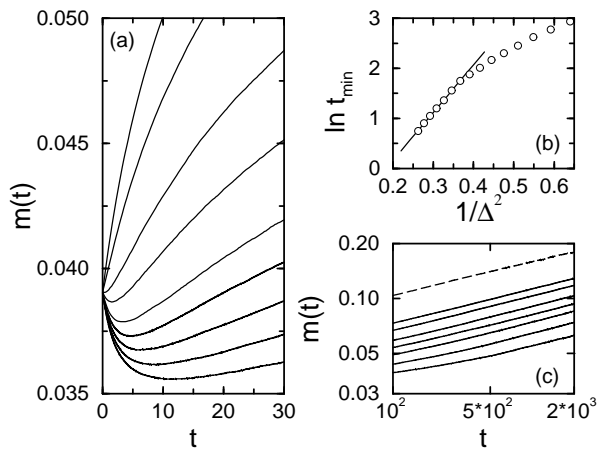}
}
\caption{(a) Magnetization vs time for various values of the variance $\Delta$: 
1.4, 1.5, 1.6, 1.7, 1.8, 2.0, 2.2, 2.6, 3.0 (from bottom to top). (b) Logarithm of $t_{min}$ as
function of $1/\Delta^2$ for $1.25 \leq \Delta \leq 1.95$. For variances $\Delta \geq \Delta_c \approx 1.65$
the data follow a straight line with slope 9.55. Deviations appear for $\Delta < \Delta_c$. Error 
bars are smaller than the size of the symbols. (c)
Magnetization vs time in the regime 2 for $\Delta =$ 1.4, 1.6, 1.8, 2.0, 2.2, 2.6, 3.0 (full lines from 
bottom to top) in a log-log plot. The dashed line is obtained when one starts from an infinite temperature initial state.
All curves yield the same slope for large $t$.
\label{fig3}}
\end{figure}

We have studied this intriguing behaviour of the magnetization in a systematic way by varying the variance
of the random fields over a large range. Some of our main findings are summarized in Figure \ref{fig3} where
we present results obtained for the initial magnetization $m_i = 0.0390625$. We checked that our conclusions
do not depend on the chosen value of $m_i$ by simulating systems with other values of the initial magnetization.

Figure \ref{fig3}a shows the temporal evolution of the magnetization at short times for various values
of the variance $\Delta$. Starting with the smallest value one observes that 
by increasing $\Delta$ the minimum gets
shallower and its position is shifted to smaller $t$. At a threshold value $\Delta_t \approx 2.0$ (the exact value
of $\Delta_t$ depends on the value of $m_i$) the minimum disappears and the magnetization increases monotonically
at early times. As mentioned before, the existence of a minimum follows from the competition of two processes,
cluster dissolution and domain growth. When $\Delta$ increases, the breaking-up length (and therefore also the
effective length of the cells) decreases, making the cluster dissolution less and less effective, up to the point
where domain growth is the only relevant mechanism, yielding an increasing magnetization immediately after the
quench.

Figure \ref{fig3}b gives a closer look at the location of the minimum. Plotting $\ln t_{min}$ as a
function of $1/\Delta^2$, we see that the points with $\Delta \in \left[ 1.65, 1.95 \right]$ lie on
a straight line, in complete agreement with the theoretically expected behaviour (\ref{tmin2}). For
values of $\Delta < 1.65$, however, deviations from the straight line behaviour are observed. Recalling that
the percolation transition takes place at $\Delta_c \approx 1.65$, we interpret these deviations
as effects on the properties of the nonequilibrium relaxation produced by the underlying 
percolation transition. The percolation point thus introduces a finite time and a finite length scale
which are the values of $t_{min}$ and $l_b$ at $\Delta_c$.

Finally, Figure \ref{fig3}c shows the time dependence of the magnetization in the intermediary 
regime 2. For any value of $\Delta$ we end up with a power-law increase with an exponent whose
value is fully compatible with the value $\theta \approx 0.19$ of the initial-slip exponent of the
two-dimensional Ising model, see (\ref{t_i}) and the dashed line in Figure \ref{fig3}c. 
It may, however, take quite a long time before reaching this simple power-law regime, as the
crossover time increases for decreasing $\Delta$.

\begin{figure}
\centerline{\epsfxsize=3.1in\epsfbox
{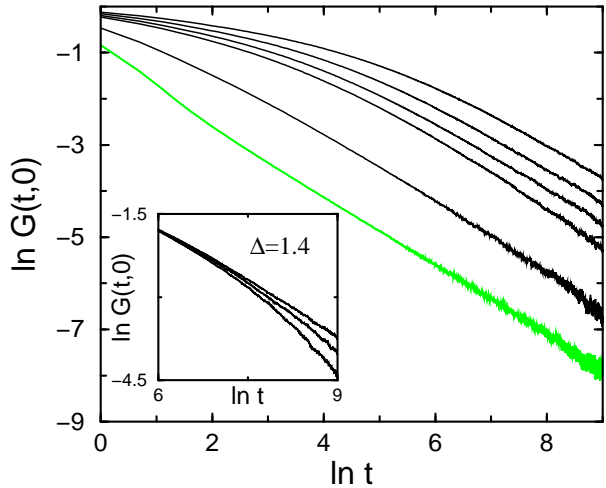}
}
\caption{Temporal evolution of the autocorrelation function $G(t,s=0)$ for different values of $\Delta$:
1.4, 1.6, 1.8, 2.0, 4.0 (from top to bottom). The grey line results from an infinite temperature initial state.
The inset shows the same quantity for $\Delta =1.4$ and systems with different linear extend $L$: 128, 200, 500 (from bottom to top).
\label{fig4}}
\end{figure}

We end the presentation of our numerical data with a discussion of the
autocorrelation function. As already mentioned, see Eq.\ 
(\ref{G_Tinf}), the autocorrelation follows a simple power law in the
long time limit when quenched from infinite temperature to the
critical point, with an exponent $\lambda/z = d/z - \theta$. As shown
in Figure \ref{fig3}c relaxation measurements yield in the regime 2 a
common, $\Delta$ independent, value of $\theta$. We therefore expect
to regain for the autocorrelation the same value for $\lambda/z$ for
any value of $\Delta$, even so the non-trivial initial state may have
a long lasting impact on the spin configurations. This is indeed what
we observe, as shown in Figure \ref{fig4}. Plotting $\ln G(t,0)$ as a
function of $\ln t$ yields at late times straight lines with a common
slope $\lambda/z \approx 0.73$. Corrections at early times are more
important for smaller values of $\Delta$, i.e. for larger effective
cell sizes in the initial state, as expected. We finally mention that
for small values of $\Delta$ strong finite-size effects appear, see
the inset of Figure \ref{fig4}, which forced us to simulate systems
containing as many as $500 \times 500$ spins.

\section{Discussion}
\label{sec:dis}
We have reported in this work an intriguing reentrance in time in
critical relaxation measurements which start from a non-trivial
initial state given by a ground state of the RFIM model. Competition
between two different mechanism, the cluster dissolution of the
compact cells and the usual domain growth, is responsible for this
novel feature in nonequilibrium critical dynamics. Note that by
varying the strength of the random field, $\Delta$, and the value of
the initial magnetization, $m_i$, one can tune the borders of the
different regimes in Figure \ref{fig1}, so that both $t_{min}$ and
$t_i$ can be macroscopic.
Interestingly, the percolation point
in the initial states, given by a critical variance of the random
fields, has an impact on the nonequilibrium dynamics by introducing
new finite time and length scales.

Finally, we note that nonequilibrium relaxation based on random fields
can be applied for other systems, too. For example for the 3d Ising
model the initial state is either ordered (for weak random fields) or
disordered (for strong random fields)\cite{natt}. Most interesting is, however,
the borderline situation when at the critical value of the strength
of the random fields the initial state is critical and described by a
random fixed point.

\acknowledgments
The authors are indebted to J-Ch. Angl\'es d'Auriac for useful discussions.
This work has been
supported by a German-Hungarian exchange program (DAAD-M\"OB), by the
Hungarian National Research Fund under grant No OTKA TO34138, TO37323,
TO48721, MO45596 and M36803, and by the Deutsche Forschungsgemeinschaft
under grant No. PL 323/2.


\begin{thebibliography}{99}

\bibitem{bray}
        A.J. Bray, Adv. Phys. {\bf 43}, 357 (1994).

\bibitem{godr}
       C. Godr\`eche and J.M. Luck, J. Phys. Condens. Matter {\bf 14},
1589 (2002).

\bibitem{ritort}
       A.\ Crisanti and F.\ Ritort, J.\ Phys.\ A {\bf 36},
R181 (2003).

\bibitem{cg04} P. Calabrese and A. Gambassi, J.\ Phys.\ A {\bf 38}, R133 (2005).

\bibitem{jss89}
        H.K. Janssen, B. Schaub, and B. Schmittmann, Z. Phys. B {\bf 73}, 539 (1989).

\bibitem{huse89}
        D. Huse, Phys. Rev. B {\bf 40}, 304 (1989).
        
\bibitem{jans92} H.K. Janssen, {\it From Phase Transition to
          Chaos}, eds. G. Gy\"{o}rgyi, I. Kondor, L. Sasv\'{a}ri and
        T.\ T\'{e}l, {\it Topics in Modern Statistical Physics} (World
        Scientific, Singapore, 1992).

\bibitem{binder} K. Binder, in {\it Phase Transitions and Critical
    Phenomena}, edited by C. Domb and J.L. Lebowitz (Academic Press,
  London, 1983), Vol. 8.
  
\bibitem{diehl} H.W. Diehl in {\it Phase Transitions and Critical
    Phenomena}, edited by C. Domb and J.L. Lebowitz (Academic Press,
  London, 1986, Vol. 10; H.W. Diehl, Int.\ J.\ Mod.\ Phys.\ B {\bf
    11}, 3503 (1997).

\bibitem{ipt}
        F. Igl\'oi, I. Peschel, and L. Turban, Adv. Phys. {\bf 42}, 683
        (1993).

\bibitem{pleim}
        M. Pleimling, J.\ Phys. A {\bf 37}, R79 (2004).

\bibitem{Berthier}
        L. Berthier, P.C.W. Holdsworth, and M. Sellitto, J. Phys. A {\bf 34}, 1805 (2001).

\bibitem{abriet}
  S. Abriet and D. Karevski, Eur. Phys. J. B {\bf 37}, 47 (2004).

\bibitem{abriet1}
        S. Abriet, PhD thesis Universit\'e de Nancy I (2004).

\bibitem{picone}
  A. Picone and M. Henkel, J. Phys. A {\bf 35} 5575 (2002).

\bibitem{aizenman}
        M. Aizenman and J. Wehr, Phys. Rev. Lett. {\bf 62}, 2503
        (1989); erratum {\bf 64}, 1311 (1990).

\bibitem{seppala}
        E.T. Sepp\"al\"a and M.J. Alava, Phys. Rev. E {\bf 63}, 066109 (2001).

\bibitem{kornyei}
  L. K\"ornyei and F. Igl\'oi (unpublished).

\bibitem{staufferaharony}
  See, for example, D. Stauffer and A. Aharony, {\it Introduction to Percolation Theory},
  (Taylor and Francis, London) (1992).

\bibitem{heiko}
   A.K. Hartmann and H. Rieger, {\it Optimization Algorithms in Physics}  (Wiley-VCH, Berlin, 2002)

\bibitem{lb}
  K. Binder, Z. Phys. B {\bf 50}, 343 (1983).

\bibitem{natt} See, for example, T. Nattermann, in {\it Spin Glasses and Random Fields}, ed. A.P. Young
(World Scientific, Singapore, 1998).

\end{thebibliography}
\end{document}